\documentclass{ieeeaccess}
\usepackage{cite}
\usepackage{amsmath,amssymb,amsfonts}
\usepackage{algorithmic}
\usepackage{graphicx,physics}
\usepackage{textcomp,bm,xcolor,soul}
\def\BibTeX{{\rm B\kern-.05em{\sc i\kern-.025em b}\kern-.08em
    T\kern-.1667em\lower.7ex\hbox{E}\kern-.125emX}}
\begin{document}

\title{Shot Optimization in Quantum Machine Learning Architectures to Accelerate Training}
\author{\uppercase{Koustubh Phalak}\authorrefmark{1}, \IEEEmembership{Student Member, IEEE},
\uppercase{and Swaroop Ghosh\authorrefmark{1}},
\IEEEmembership{Senior Member, IEEE}}
\address[1]{Computer Science and Engineering Department, Pennsylvania State University, PA 16802}
\tfootnote{The work is supported by NSF (CNS -1722557, CNS-2129675, CCF-2210963, CCF-1718474, OIA-2040667, DGE-1723687, DGE-1821766, and DGE-2113839) and seed grants from Penn State ICDS.}

\markboth
{Phalak \headeretal: Shot Optimization in Quantum Machine Learning Architectures to Accelerate Training}
{Phalak \headeretal: Shot Optimization in Quantum Machine Learning Architectures to Accelerate Training}


\begin{abstract}
Quantum Machine Learning (QML) has recently emerged as a rapidly growing domain as an intersection of Quantum Computing (QC) and Machine Learning (ML) fields. Hybrid quantum-classical models have demonstrated exponential speedups in various machine learning tasks compared to their classical counterparts. On one hand, training of QML models on real hardware remains a challenge due to long wait queue and the access cost. On the other hand, simulation-based training is not scalable to large QML models due to exponentially growing simulation time. 
Since the measurement operation converts quantum information to classical binary data, the quantum circuit is executed multiple times (called shots) to obtain the basis state probabilities or qubit expectation values. 
Higher number of shots worsen the training time of QML models on real hardware and the access cost. Higher number of shots also increase the simulation-based training time. 
In this paper, we propose shot optimization method for QML models at the expense of minimal impact on model performance. We use classification task as a test case for MNIST and FMNIST datasets using a hybrid quantum-classical QML model. 
First, we sweep the number of shots for short and full versions of the dataset. We observe that training the full version provides 5-6\% higher testing accuracy than short version of dataset with up to 10X higher number of shots for training. Therefore, one can reduce the dataset size to accelerate the training time. 
Next, we propose adaptive shot allocation on short version dataset to optimize the number of shots over training epochs and evaluate the impact on classification accuracy. We use a (a) linear function where the number of shots reduce linearly with epochs, 
and (b) step function where the number of shots reduce in step with epochs. 
We note around 0.01 increase in loss and maximum $\bm{\sim}$4\% (1\%) reduction in testing accuracy for reduction in shots by up to 100X (10X) for linear (step) shot function compared to conventional constant shot function for MNIST dataset, and 0.05 increase in loss and $\bm{\sim}$5-7\% (5-7\%) reduction in testing accuracy with similar reduction in shots using linear (step) shot function on FMNIST dataset. For comparison, we also use the proposed shot optimization methods to perform ground state energy estimation of different molecules and observe that step function gives the best and most stable ground state energy prediction at 1000X less number of shots. \color{black}
\end{abstract}


\titlepgskip=-15pt

\maketitle

\section{Introduction}
\label{sec:introduction}
\PARstart{\raisebox{3pt}{Q}}{ML} has gained a lot of traction since the past decade with recent advancements in quantum computers \cite{gill2022quantum}. In QML models, classical data is converted to quantum Hilbert space \cite{schuld2019quantum}, quantum operations are performed using parametric quantum gates to change the state of the qubits and classical measurement like expectation value or probability of computational basis states are performed based on some observables like Pauli X/Y/Z operators. Quantum Machine Learning (QML) models such as trainable hybrid quantum-classical models \cite{biamonte2017quantum} combine Machine Learning (ML) methods with Quantum Computing (QC) to show exponential speedup with lesser number of trainable parameters \cite{caro2022generalization}. 
Similar quantum model performance as classical model is shown by \cite{schuld2020circuit,ai2022decompositional,di2022dawn} with up to exponential reduction in model parameters. \cite{rebentrost2014quantum, li2015experimental,lloyd2018quantum} theoretically demonstrate exponential speedup in runtime while \cite{chen2022quantum} show runtime reduction practically via simulation. While QML algorithms have shown the promising quantum advantage theoretically, their power on modern Noisy Intermediate-Scale Quantum (NISQ) computers \cite{riste2017demonstration} are currently limited by noise. As quantum computers become more fault tolerant, the quantum advantage of QML is expected to become practically realizable. Another potential hurdle for QML algorithms right now are the training times. Publicly available quantum hardware have long queues accommodated by circuits from various users from all around the world. Therefore, a user's quantum circuit may take as long as $\sim$ 9-10 hours just to reach the top of the queue. For iterative quantum circuits like Parametric Quantum Circuits (PQC), this may not even be practically possible to train on real hardware. So, users generally employ quantum simulators with or without noise to train QML models. Quantum simulators are fast due to lack of queues and can reduce training time of iterative circuits from months/days to minutes. 
However, with increasing size of quantum circuits the simulation based training will become exponentially slower motivating hardware-based training.

Other than long wait queue, NISQ computers are costly to access, especially the Quantum Processing Units (QPUs) with large number of qubits. For example, 27-qubit Falcon processors by IBM cost \$1.60 per runtime second \cite{ibmcost}, IonQ have quoted their costs at \$0.0003 for 1 single qubit gate, \$0.003 for 1 two qubit gate, and \$0.00165 per shot of quantum execution \cite{ionqcost}, Rigetti has Aspen-11 and Aspen-M quantum annealers priced at \$0.3 per task and \$0.00035 per shot \cite{rigetticost}. More pricing data on other quantum hardware from different companies like D-Wave, Xanadu and OQC can be found in \cite{rigetticost}. From these pricing data, it is evident that a QML model with relatively large depth and shots can incur very high cost which can further exacerbate for a training scenario with large dataset. 

In this paper, we propose an optimization method to reduce the number of shots used per data point to accelerate training time (which will also reduce the training cost). We use MNIST and FMNIST datasets \color{black} and perform sweep of shots on a shortened and full version of the dataset. 
Later, we propose two methods to adaptively change the number of shots, one which reduces the number of shots linearly as a function of number of epochs, and second which is a step function where the number of shots reduces by 100 every 10 epochs. 
We also use the proposed linear and step functions to perform ground state energy estimation of different molecules such as H$_2$, He$_2^{+}$, LiH, NH$_3$ and BeH$_2$.\color{black}

The structure of the paper is as follows: in Section II we introduce some relevant background and related work, in Section III we present the QML model used for training the MNIST and FMNIST datasets \color{black} for classification and describe the results and analysis, in Section IV we present the adaptive shot allocation methods, the results and a comparative analysis with existing methods. We also present the ground state energy estimation results using our proposed shot allocation methods in this section. \color{black} Finally, in Section V we conclude the paper.

\section{Background and Related Works}
In this section, we provide background on quantum computing and related works on shot optimization. 
\subsection{Background}
\paragraph{\textbf{Qubits}} Quantum bits (or qubits) are quantum equivalent of classical bits. They are fundamental units of a quantum computer. While classical bits only have either 0 or 1 values, a qubit can have a range of values, which can be in general represented as a quantum state $\ket{\psi}=$
$\big[\begin{smallmatrix}
  \alpha\\
  \beta
\end{smallmatrix}\big]$, where $|\alpha^2|$ represents the probability of qubit being measured to 0, and $|\beta^2|$ represents the probability of qubit being measured to 1. A qubit has two special states $\ket{0}$ with $\alpha=1,\beta=0$ and $\ket{1}$ with $\alpha=0,\beta=1$, these states are called computational basis states. A more general version of the qubit is the qudit, which consists of d-computational basis states. Many technologies like superconducting qubits, trapped ion qubits, neutral atom qubits, quantum annealers have been implemented to realize physical qubits on quantum systems.

\paragraph{\textbf{Quantum Gates}} A quantum gate is a unitary matrix operation performed on either a single or multiple qubits to change the state (s). 
Some commonly used single qubit gates are bit-flip gate, Hadamard gate, Pauli X/Y/Z gates, RX/RY/RZ rotation gates, identity gate, T gate, S gate and multi qubit gates include CNOT gate, SWAP gate, Controlled SWAP gate, Toffoli gate, iToffoli gate, Peres gate. Every quantum computer implements native gate set so every non-native gate in the quantum circuit has to be decomposed into the native gate set. 

\paragraph{\textbf{Quantum Circuit}} Quantum circuit is a program with ordered sequence of quantum gates. At the end of every quantum circuit, there is measurement operation that collapses the quantum state of the qubit classically. The measurement can be expectation value of an operator like Pauli X/Y/Z or probability values of computational basis states. 

\paragraph{\textbf{Shots}} Shots/trials refer to the number of repeated quantum executions of a quantum circuit. 
Usually, multiple shots are used to compute expectation values and probability values. Higher the number of shots, more accurate will be the output. Modern NISQ computers have usage pricing based on number of gates and number of shots. For number of shots, the price varies from \$0.00019 per shot all the way to \$0.01 per shot  \cite{rigetticost}.

\paragraph{\textbf{Parametric Quantum Circuit (PQC)}} A Parametric Quantum Circuit is a trainable quantum circuit similar to classical ML model. PQCs consist of parametric quantum gates like RX, RY and RZ which have trainable angle parameter $\theta$ and entangling gates like CNOT and controlled Pauli/rotation gates that improve the fidelity of computation. At the end, classical measurement is performed either of probability values or qubit expectation values using measurement gates to get the desired output. 

\paragraph{\textbf{Quantum Neural Network (QNN)}} Quantum Neural Networks are quantum equivalent of classical neural networks (NN) and are considered as an expansion on Deutsch's model of quantum computational networks \cite{deutsch1989quantum}. Data is first encoded into quantum Hilbert space using embedding methods like amplitude, angle or basis embedding, which is then followed by PQC and finally measurement operation. The PQCs are optimized using classical ML optimization methods like SGD, Adam, Adagrad, etc. QNNs can also have hybrid quantum-classical architectures consisting of classical layers and PQCs.

\color{black}

\subsection{Related Works}
Closely related work includes Coupled Adaptive Number of Shots (CANS) \cite{kubler2020adaptive, gu2021adaptive} and Random Operator Sampling for Adaptive Learning with Individual Number of shots (Rosalin) \cite{arrasmith2020operator} methods. The CANS method is a modification of Coupled Adaptive Batch Size (CABS) method \cite{balles2016coupling} which adaptively selects the batch size by maximizing the lower bound on the the expected gain per computational cost. The CANS method modifies the CABS method by adaptively selecting the number of shots with the same criteria of maximizing lower bound on the expected gain per computational cost. CANS method has two variants, first is individual-CANS (iCANS) \cite{kubler2020adaptive} which selects different number of shots $s_i$ for each individual partial derivative $g_i$ and picks the maximum shot value $s_{arg,max} = max(s_i) \forall i$ and putting a soft cap on the maximum shot value $s_{max}$ so that $s_{arg,max}$ does not exceed this value, and second is global-CANS (g-CANS) \cite{gu2021adaptive} which determines a single shot value $s$ based on the full magnitude of gradient $|g|$. Rosalin method \cite{arrasmith2020operator} augments iCANS method with a weight sampling method to make the estimator of cost function unbiased. 

iCANS optimizer is used for solving Heisenberg spin chain Variational Quantum Eigensolver (VQE), while gCANS optimizer is used to solve chemical configuration problems and also find ground state of an Ising model with different number of spins. Rosalin optimizer is used to compute the ground states of different molecules. 

In theory, the above works can also be extended to optimize the number of shots for general QML workloads. However, such efforts have not been pursued in literature. Furthermore, gradient-based shot optimization may incur stability issues for general QML workloads which may not have well-behaved gradients similar to molecular problems. To address these concerns, we propose a simpler alternative such as, adaptive shot allocation which removes the dependency of shot calculation on gradients and instead computes the number of shots as a function of number of epochs/iterations \color{black} completed.

\begin{figure}[t]
    \centering
    \includegraphics[width=0.87\linewidth]{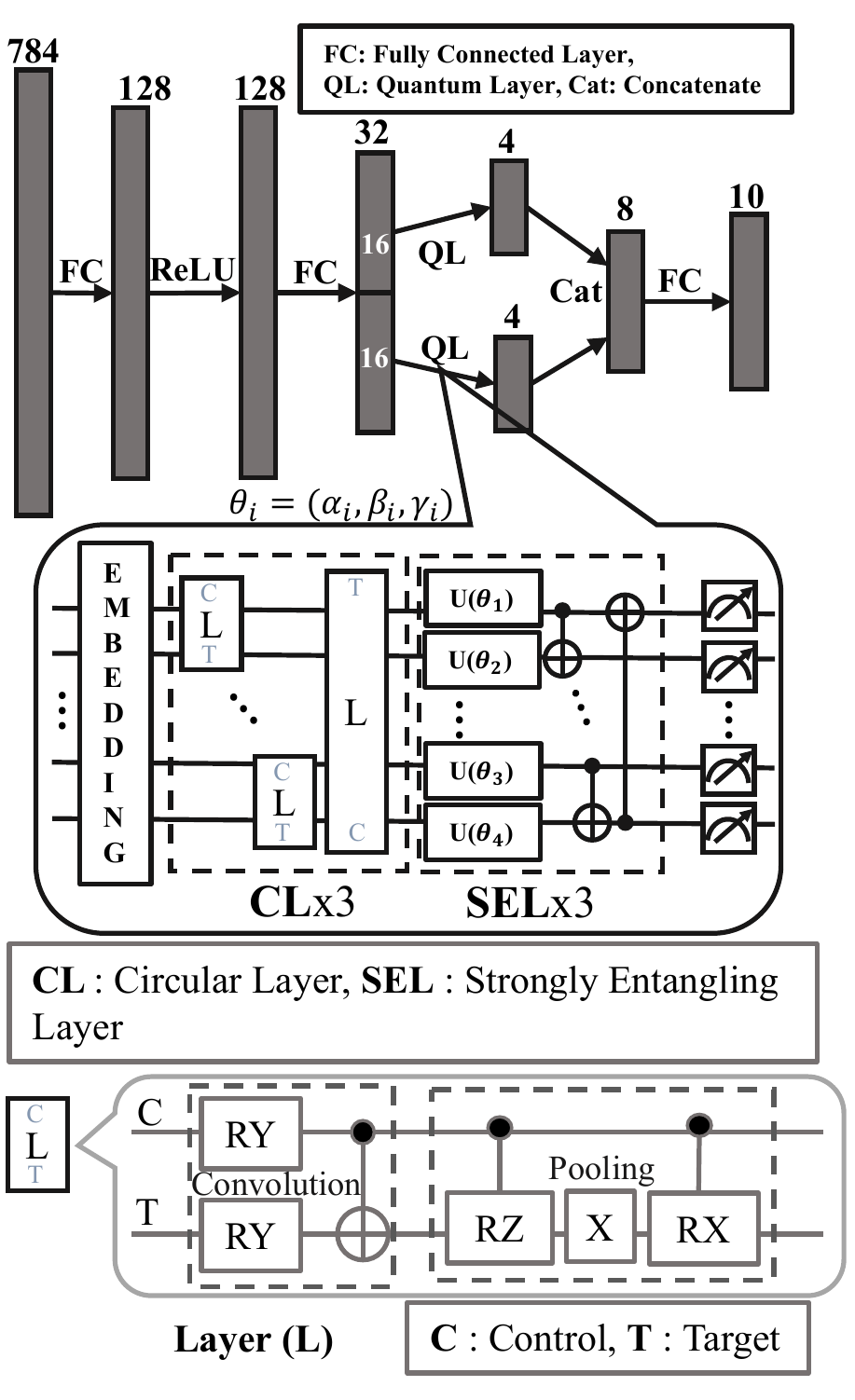}
    \caption{Hybrid quantum-classical model used for training MNIST and FMNIST datasets for classification.}
    \label{fig:qml_model}
    \vspace{-6mm}
\end{figure}

\begin{figure*}[t]
    \centering
    \includegraphics[width=0.9\linewidth]{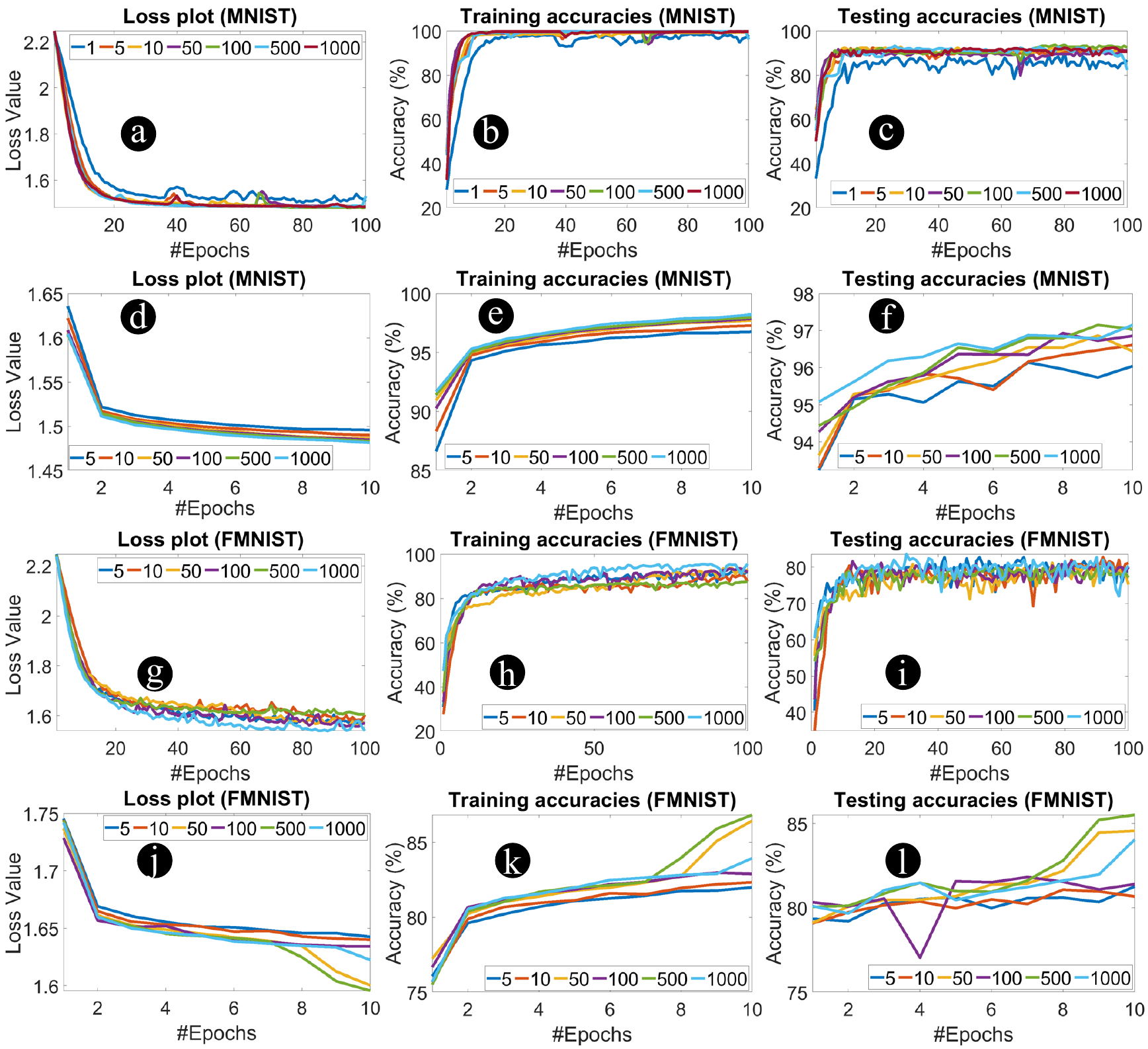}
    \vspace{-3mm}
    \caption{Sweep plots for MNIST (FMNIST): number of shots (a),(b),(c) ((g),(h),(i)) training \& testing loss plots, training and testing accuracies respectively for short version, and (d),(e),(f) ((j),(k),(l)) training \& testing loss plots, training accuracies and testing accuracies respectively for full dataset.}
    \label{fig:sweep_results}
    \vspace{-4mm}
\end{figure*}

\begin{figure*}[t]
    \centering
    \includegraphics[width=0.9\linewidth]{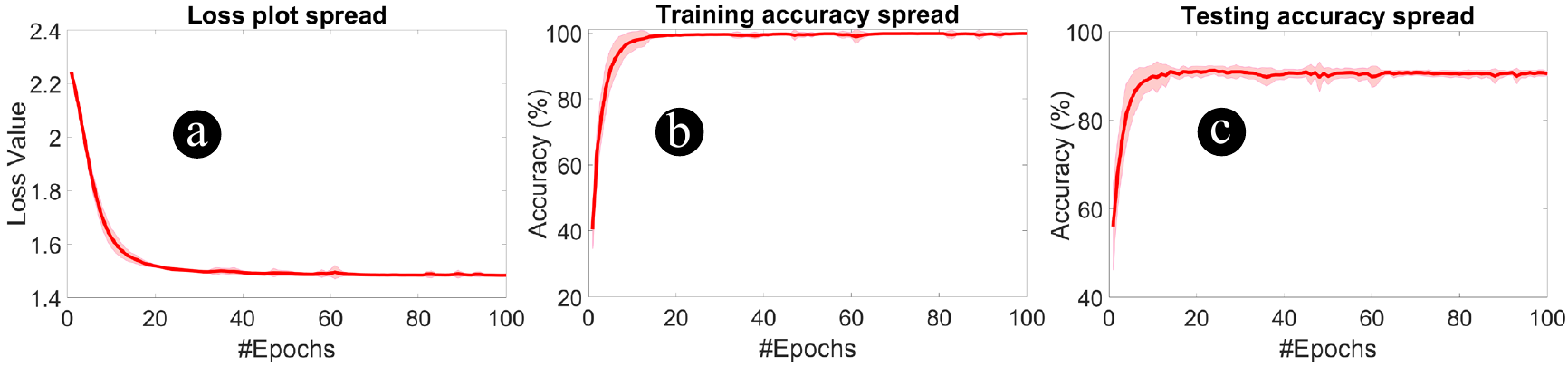}
    \vspace{-3mm}
    \caption{Effect of different initializations on training. 15 different starting initializations are performed with short version of MNIST dataset, 1000 shots in each case and (a) loss plot, (b) training accuracy and (c) testing accuracy spreads are plotted.}
    \label{fig:init_spread}
    \vspace{-6mm}
\end{figure*}

\begin{figure*}[t]
    \centering
    \includegraphics[width=\linewidth]{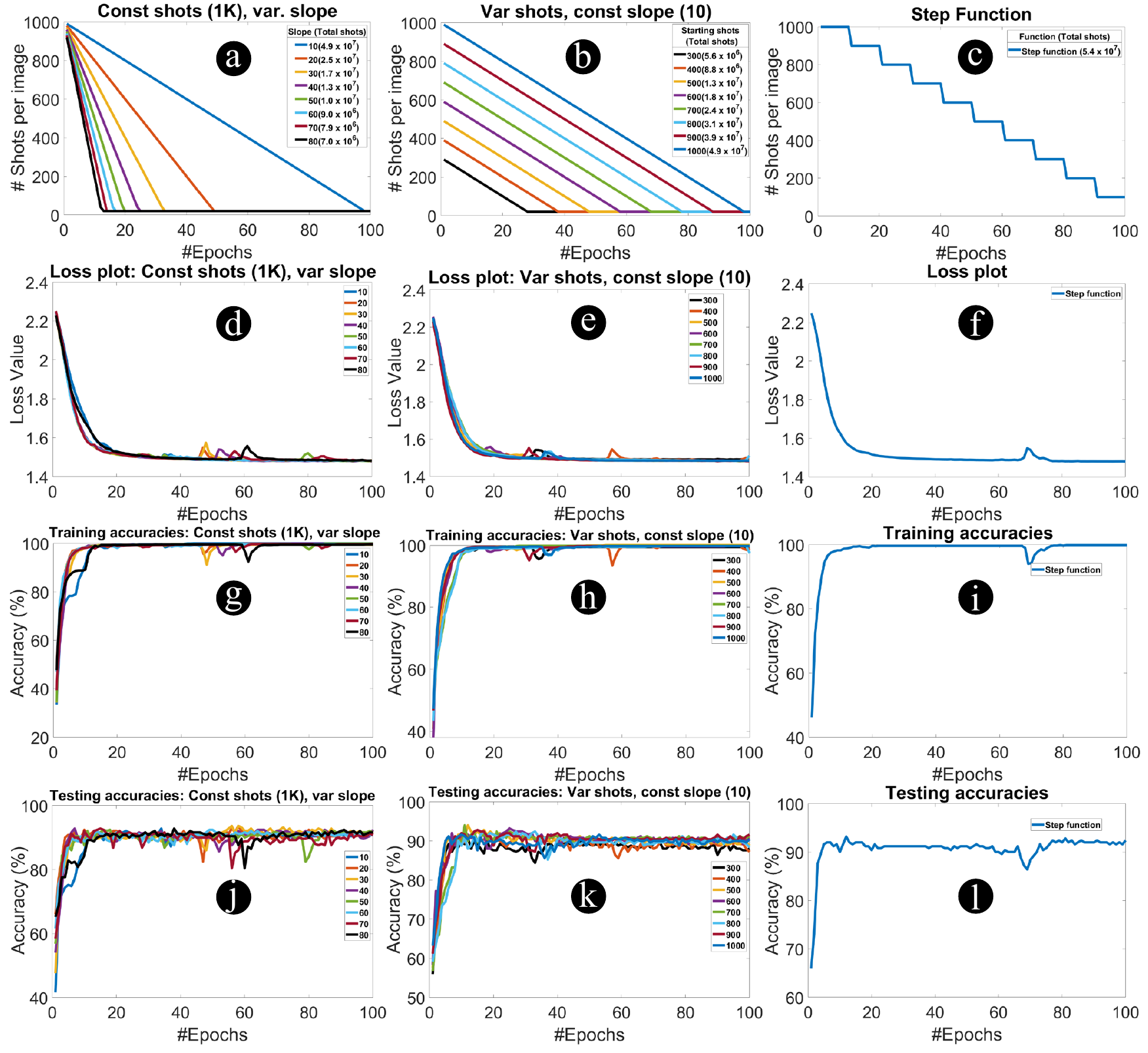}
    \vspace{-8mm}
    \caption{MNIST dataset plots for adaptive shot allocation. Shot function, loss plots, training and testing accuracies, respectively for linear function ($s_t = max(20, s_{start} - slope * t)$) with constant starting shots ($s_{start}$) (a),(d),(g),(j), linear function with constant slope ($slope$) (b),(e),(h),(k) and step function ($s_t = 1000 - 100*\lfloor\frac{t}{10}\rfloor$) (c),(f),(i),(l). In each of the shot function plots (a),(b),(c), we mention the total number of shots used for training for each case.}
    \label{fig:adap_shot_plot}
    \vspace{-5mm}
\end{figure*}

\section{Analysis of hybrid quantum-classical model}
\subsection{Basic QML setup}
We use a hybrid QML model consisting of classical and quantum layers to classify the images in MNIST dataset. For classical layers, we use fully connected layers and ReLU activation function and for quantum layers, we use two 4-qubit Parametric Quantum Circuits (PQC). The PQC uses amplitude embedding to map 16 features to 4 qubits, followed by 3 circular layers consisting of convolution ansatz 1 and pooling ansatz from \cite{hur2022quantum}(shown in lower bubble of Fig. \ref{fig:qml_model}) \color{black} and three strongly entangling layers \cite{stronglayers}. We pick ansatz 1 from \cite{hur2022quantum} because it has lowest circuit depth, which is useful in providing better fidelity of computation during noisy hardware runs. We apply strongly entangling layers to increase overall entanglement in the circuit. According to \cite{cai2015entanglement}, having strongly entangled qubits can provide exponential speedup for classification problems. \color{black} All MNIST images are first flattened from 28*28 size to a vector of size 784. This size is reduced to 128 via a fully connected layer. A ReLU activation function is used after the fully connected layer to improve convergence. This output is then applied to another fully connected layer which further reduces the vector size to 32, which is split to two vectors of size 16 which are sent to two different quantum layers having their own set of parameters. Note that we use two 4-qubit PQCs because we get better final classification accuracy as compared to using a single 5-qubit PQC that embeds 32 features on 5 qubits. \color{black} Both the quantum layers compute PauliZ expectation values of the qubits to give 4 dimensional vectors. These 4 dimensional vectors are combined together to a vector of size 8. A final fully connected layer is used to scale the output dimension to size 10 which is used to compute multi-cross entropy loss for 10 classes of digits. The full QML model along with the structure of the 4 qubit PQC is shown in Fig. \ref{fig:qml_model}. We initialize the parameters of our model from a normal distribution with mean 0 and standard deviation of 1 such that every time the initialization values of the parameters are different.

We first perform sweep on the number of shots during training of this QML model for two different sizes of MNIST and FMNIST datasets. \color{black} The first dataset contains 1000 training images (100 randomly chosen per class) and 250 testing images (25 randomly chosen per class) and is trained for 100 epochs, and the second dataset contains the complete training (60,000 images) and testing (10,000 images) set of MNIST, FMNIST datasets \color{black} and is trained for 10 epochs. We analyze the impact of parameter initialization values and the proposed adaptive shot allocation on the shorter version of the dataset.

\subsection{Impact of the number of shots and dataset size}
We sweep the number of shots from 1 to 1000 to study the impact on training and testing accuracies and losses. We use both the short and full versions of MNIST and FMNIST datasets \color{black} to estimate the impact of dataset size in this study as well. 
Note that the shots numbers correspond to per image. We plot the training and testing loss plots, training accuracies and testing accuracies for short version of the datasets in Fig. \ref{fig:sweep_results}(a), (b) and (c) (MNIST), and (g), (h), and (i) (FMNIST) \color{black} respectively and for full dataset in Fig. \ref{fig:sweep_results}(d), (e) and (f) (MNIST), and (j), (k) and (l) (FMNIST) \color{black} respectively.

The plots show a general trend of lower performance with reduced shots. While this trend is not so clear for short version of both the datasets (Fig. \ref{fig:sweep_results}(a),(b) MNIST and (g),(h) FMNIST\color{black}), the trend is more clear for the full dataset (Fig. \ref{fig:sweep_results}(d),(e)). This trend is expected because a single shot refers to single quantum execution of the PQC model, and higher the number of shots, more accurate will be the expectation value calculated for the qubits. Next, we observe that the training accuracy for short version of the dataset converges to 99-100\% (MNIST) and 93-95\% (FMNIST)\color{black}, and for the full dataset to around 97-98\% (MNIST) and 80-85\% (FMNIST). We also observe testing accuracy for short version of the dataset at around 90-91\%  (MNIST) and 75-80\% (FMNIST) and for full dataset at around 96-97\% (MNIST) and 80-85\% (FMNIST). Analyzing these observations, we note that (i) The short version of the datasets have higher training accuracy and lower testing accuracy compared to their full dataset counterparts. This implies that the full dataset is able to provide better generalization with higher testing accuracy and lower training accuracy. (ii) MNIST dataset performs better overall compared to FMNIST dataset. This is expected because FMNIST is a more complex dataset with multiple classes having lot of similarities. For example, as can be seen in Table 3 in \cite{xiao2017fashion} multiple images of Top, Pullover, Dress, Coat, and Shirt look similar. There is also a similar resemblance between Sandal, Sneaker and Ankle Boots images. These similarities make it harder for the model to identify the correct class of the image.\color{black}


For MNIST dataset as we go from full dataset to the short version, we note 0.02 increase in loss and $\sim$1\% reduction in testing accuracy for reduction in shots from $10^8$ to $0.5*10^6$. For FMNIST dataset we note 0.05 increase in loss for similar reduction in shots. \color{black} Finally, we observe that the training curves for full dataset are smooth, while the short version have bumps. For example, in Fig. \ref{fig:sweep_results}(a) and (b) at around $40^{th}$ epoch we observe an increase in loss and decrease in training accuracy in the form of bumps for certain shot values. For FMNIST dataset we observe that the bumps are smaller but more spikier. \color{black} These bumps are formed due to small size of the dataset. If the size of the dataset is increased, the frequency and size of bumps will reduce. In the remaining analysis we use the reduced datasets for adaptive shot allocation using linear and step shot functions.
\color{black}
\subsection{Impact of parameter initialization}
We initialize the model parameters from a normal distribution with mean 0 and standard deviation 1. A good initialization can accelerate training by starting from a point that is close to global minima on the loss landscape \cite{narkhede2022review}. We choose 15 such starting initializations and train the model on MNIST dataset \color{black} for each case. The loss, training accuracy and testing accuracy for each case and the spread curve for each case is shown in Fig. \ref{fig:init_spread} (a), (b) and (c), respectively. From these plots, our general observations are (i) the variation (light pink color) for all the three curves are biggest in the range of 0-20 epochs (loss:$\pm 0.04$, training accuracy: $\pm7.5\%$, testing accuracy: $\pm9.77\%$), (ii) beyond 20 epochs the variation for loss and training accuracy curves becomes negligible while for testing accuracy it is small but non-negligible in the range of 20-60 epochs ($\pm3.3\%$) and becomes negligible beyond 60 epochs, and (iii) by $100^{th}$ epoch, all the curves have a negligible variation which implies a more deterministic solution. Therefore, we conclude that as long as we initialize the parameters from the normal distribution, we can expect a deterministic final solution. 
We roughly observe a fast convergence of training at around $20^{th}$ epoch due to good parameter initialization. We can also safely extrapolate these observations to other different shot values as the the behavior of parameter initializations is independent of the number of shots used.  \color{black}

\begin{figure*}[t]
    \centering
    \includegraphics[width=\linewidth]{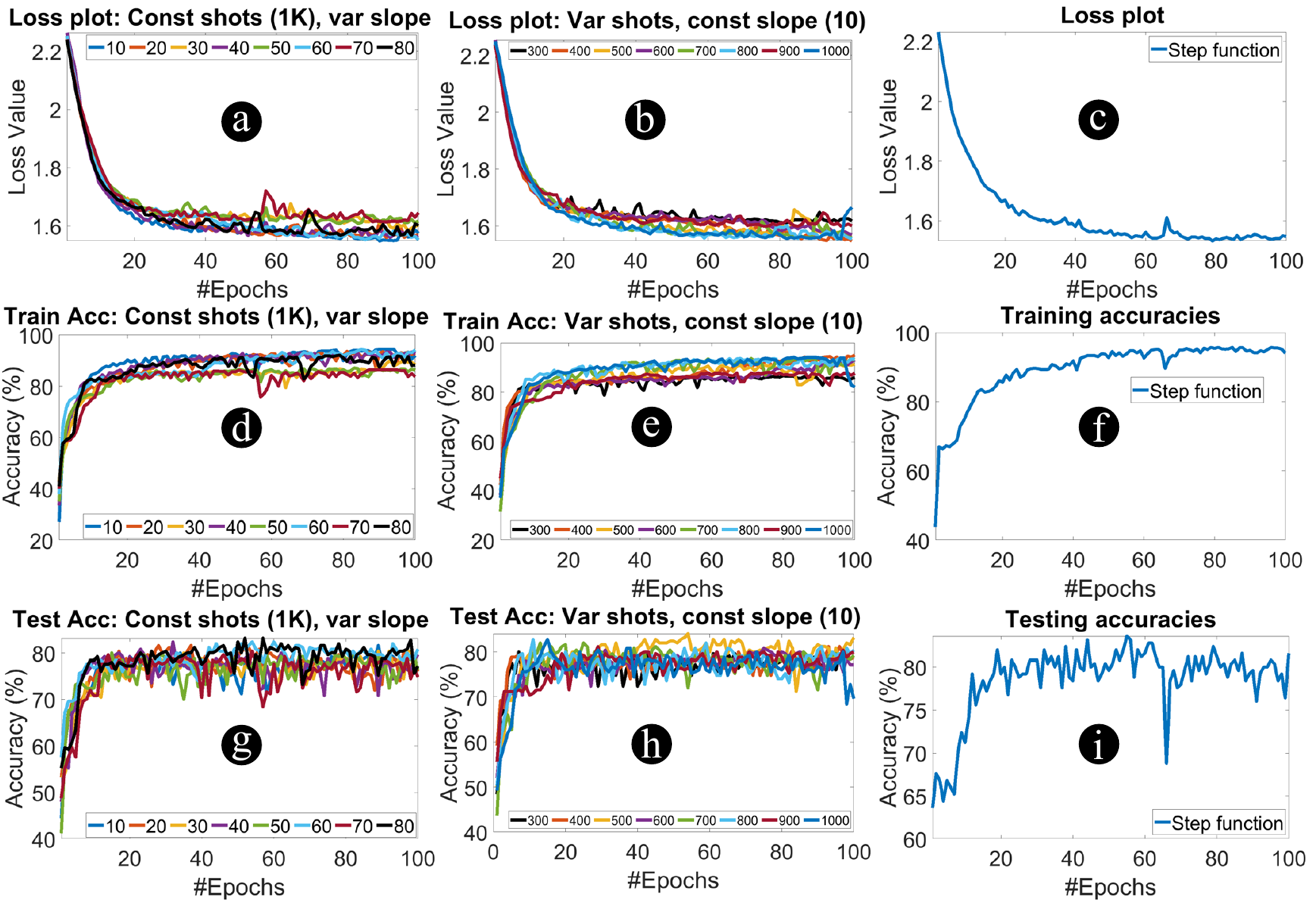}
    \vspace{-8mm}
    \caption{FMNIST dataset adaptive shot allocation plots. loss plots, training and testing accuracies, respectively for linear function (($s_t = max(20, s_{start} - slope * t)$)) with constant starting shots ($s_{start}$) (a),(d),(g), linear function with constant slope ($slope$) (b),(e),(h), and step function ($s_t = 1000 - 100*\lfloor\frac{t}{10}\rfloor$) (c),(f),(i).}
    \label{fig:adap_shot_plot_2}
\end{figure*}

\begin{table*}[t]
    \centering
    \includegraphics[width=\linewidth]{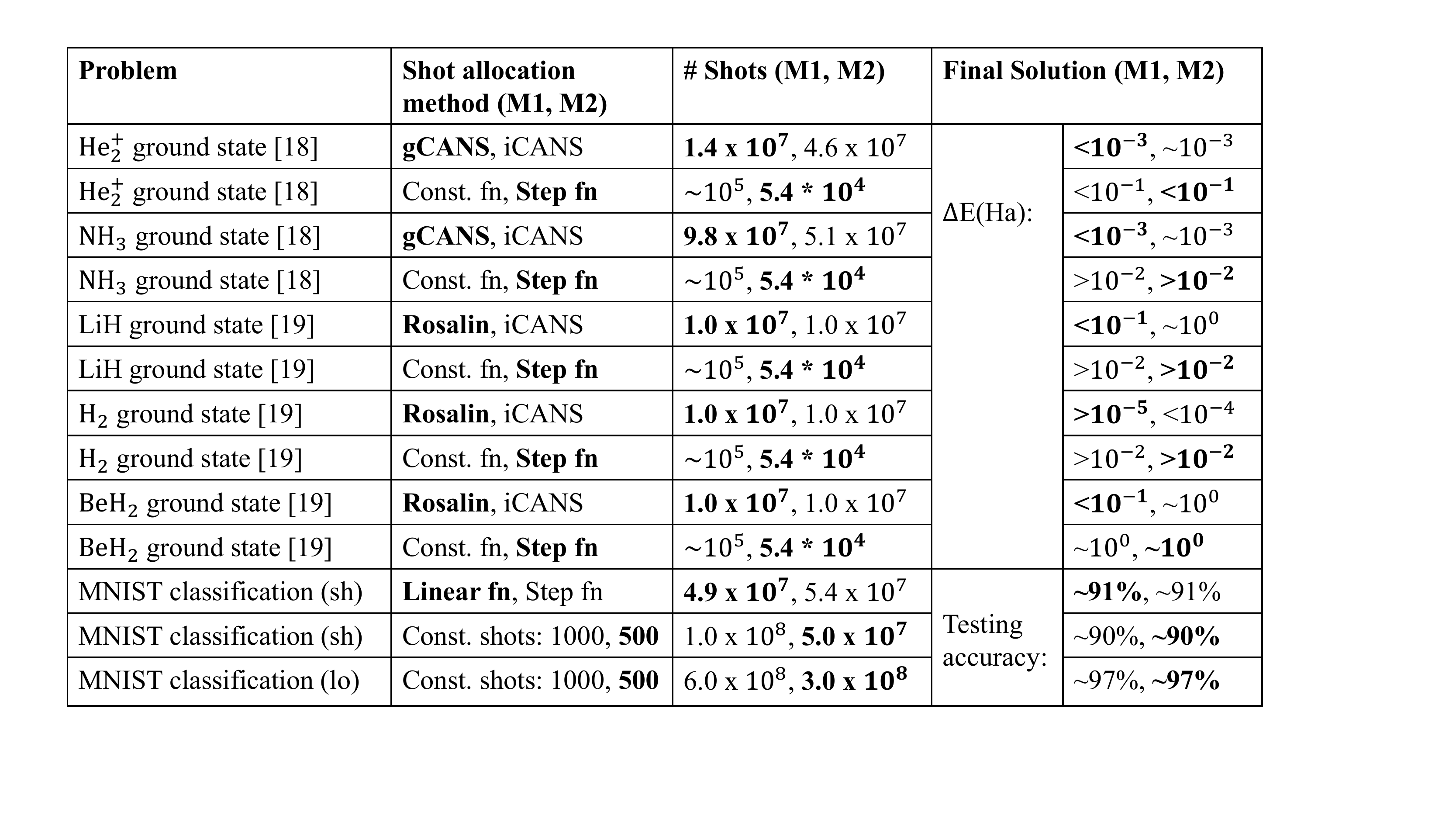}
    \caption{Comparison of number of shots used and solution quality for various shot allocation methods for different problems. Note that sh: short version, lo: full MNIST dataset (long).}
    \label{tab:table_of_comparison}
    \vspace{-8mm}
\end{table*}

\begin{figure*}[t]
    \centering
    \includegraphics[width=\linewidth]{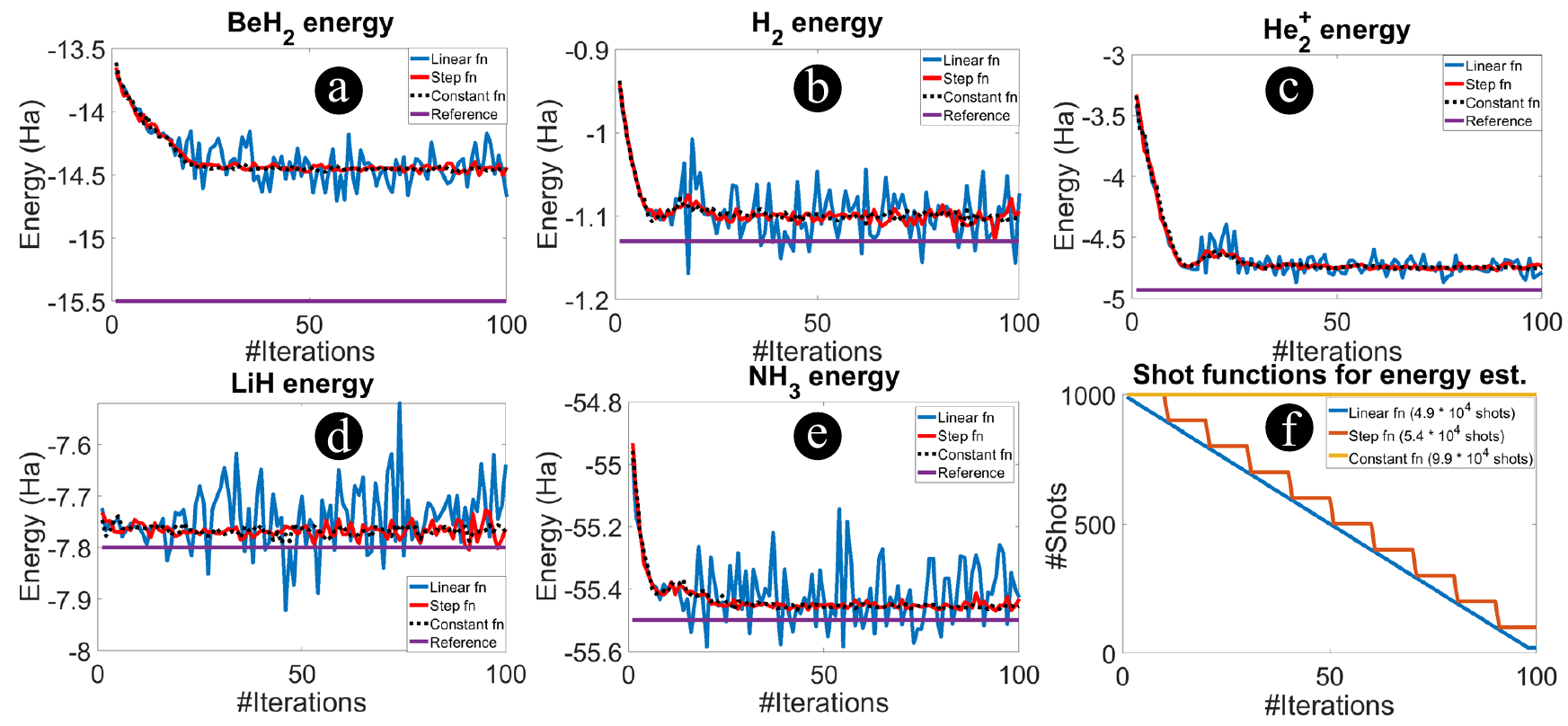}
    \caption{Ground state energy prediction plots for (a)BeH$_2$, (b)H$_2$, (c)He$_2^{+}$, (d)LiH and (e)NH$_3$ molecules. We also show the shot functions in (f).}
    \label{fig:ground_state_plot}
\end{figure*}

\section{Proposed adaptive shot allocation}
\subsection{Basic idea}
Based on the analysis from previous Section, it is evident that higher number of shots may not be needed for QML tasks. We propose two methods to adaptively reduce the number of shots over training epochs: linear function method and step function method. We represent the general linear function as $s_t = s_{start} - slope * t$, where $s_t$ represents the shots being used in $t^{th}$ epoch per image, $s_{start}$ represents the starting number of shots per image, and $slope$ represents the rate at which $s_t$ reduces. We perform a sweep on $s_{start}$ (with constant $slope=10$) and $slope$ (with constant $s_{start}=1000$), where we vary $s_{start}$ from 300 to 1000 shots in steps of 100 and $slope$ from 10 to 80 in steps of 10. We maintain a lower bound of 20 shots for each of the sweep parameters ($s_t = max(20, s_{start} - slope * t)$) to ensure that expectation value calculation is being done on sufficient number of classically measured outputs. We denote the step function using the formula $s_t = 1000 - 100*\lfloor\frac{t}{10}\rfloor$, where $\lfloor.\rfloor$ is the floor function. After every 10 epochs, we reduce the number of shots used per image by 100. We start with 1000 shots and continue this process for 100 epochs. We show the linear and step shot functions in Fig.  \ref{fig:adap_shot_plot} (a), (b) (linear) and (c) (step) respectively.\color{black}

The rationale for choosing linear and step functions is to have higher shots at the start of the training and then gradually reduce the number of shots with training epochs. The reason for doing this is (i) to ensure that the model learns well in the starting epochs of training. Note that higher number of shots implies more accurate expectation values or probability values. Having a more accurate result at the start is preferable so that the model can learn quickly. (ii) Once the model has learnt sufficiently i.e., reached/reaching saturation, using same number of shots would be wastage of computing resources and training time. Therefore, we reduce the number of shots as a function of epochs to save this wastage. We would want to put a lower bound of the minimum number of shots used to ensure the computation of measurement is accurate enough and stable. One may also wonder how fast or how slow should the number of shots be reduced. In general, we do not want the shots to reduce very fast as it will affect learning in the initial epochs, or too slow as it will lead to wastage of shots.\color{black}

As we show in the next subsection, we perform a sweep on the slope and starting number of shots for the linear function and show the results for optimal slope and starting shots.\color{black}

\subsection{Results}
We plot the shot function curves and training results for both the adaptive shot allocation methods in Fig. \ref{fig:adap_shot_plot} (MNIST) and Fig. \ref{fig:adap_shot_plot_2} (FMNIST). For MNIST dataset, \color{black} we observe that $s_{start}$ has more effect on testing accuracy compared to $slope$. Sweep on $slope$ at constant $s_{start} = 1000$ gives around 90-91\% testing accuracy, while sweep on $s_{start}$ with constant $slope=10$ gives more variation in testing accuracy around 87-91\%. We observe that a lower $s_{start}$ gives lower accuracy and more unstable training with bumps in the curves compared to higher $s_{start}$. We get stable training as long as $s_{start}$ is high enough (1000 shots in this case) regardless of $slope$. For all cases, we obtain 99-100\% training accuracy. We observe a minimum of $5.6 * 10^{6}$ total shots for $s_{start}=300$, $slope=10$ and maximum of $4.9 * 10^{7}$ total shots for $s_{start}=1000$, $slope=10$ for training (excluding shots used for testing).
We roughly note 0.01 increase in loss and $\sim$4\% maximum reduction in testing accuracy for reduction in shots from $4.9*10^7$ to $5.6*10^6$. \color{black} For step function, we use a total of $5.4*10^{7}$ shots for training and observe 99-100\% training accuracy and 90-91\% testing accuracy. For both the linear and step function curves, we observe bumps in the plots similar to plots of shots sweep for short version of MNIST dataset. Based on these findings, we can conclude that linear function is better than the step function as it provides similar testing accuracy (90-91\%, given $s_{start}$ is high enough) while being conservative in terms of total number of shots used for training.

For FMNIST dataset, we observe a similar trend for training curves. For constant shots, lower slope gives better training accuracy and for constant slope, higher starting shots give better training accuracy. Overall, we observe 82-92\% range of training accuracy and 70-80\% testing accuracy. However, this trend is not well-demarcated for testing accuracy curves. For example, we observe that slope of 80 (78-80\%) gives higher testing accuracy compared to slope of 50 (74-76\%), and 500 starting shots gives better testing accuracy (80-82\%) compared to 1000 starting shots (76-78\%). This implies that for a more complex dataset like FMNIST, starting with smaller number of shots with higher slope may be preferable. A possible reason for this behavior is that lower number of shots for training leads to less overfitting on training set and better generalization on testing set.\color{black}

\subsection{Comparative analysis with existing approaches}
We compare the total number of shots consumed in proposed methods with previous works in Table \ref{tab:table_of_comparison}. We depict usage for two different problems being solved namely, ground state energy prediction of different compounds (previous work) and classification of MNIST images (this work). Each row shows two methods M1 and M2, the total number of shots used, and the final solution quality. For the case of ground state energy prediction problem the final solution is the average energy above ground state ($\Delta$E in Hartree) and for MNIST classification it is final testing accuracy. He$_{2}^{+}$ and NH$_3$ ground state energies are found using gCANS and iCANS methods, and for both cases gCANS gives better solution than iCANS. However, for He$_{2}^{+}$ problem, gCANS consumes lesser shots than iCANS while for NH$_3$ problem it consumes higher shots. Rosalin and iCANS methods are used to find the ground state energies of LiH, H$_2$ and BeH$_2$ compounds where both Rosalin and iCANS use $10^7$ shots and Rosalin gives better ground state energy. For MNIST classification, we compare linear function ($s_{start}=1000$, $slope=10$) with step function (for starting shots of 1000) and fixed 1000 shots with fixed 500 shots for both short version and full dataset. Linear function and step function both give similar testing accuracy in the range of 90-91\% but linear function is better overall as it consumes lesser shots. For constant shot sweeps, we once again roughly observe 90-91\% testing accuracy for both 1000 and 500 shots for short version dataset and around 97\% testing accuracy for full dataset. So 500 shots is better as it consumes lesser shots overall while giving similar accuracy. Overall, linear function is the most conservative for MNIST dataset giving a maximum of $4.9*10^7$ total shots. These shots are higher than gCANS for He$_{2}^{+}$ case but lower for NH$_3$ case, higher than Rosalin for all LiH, H$_2$ and BeH$_2$ and also higher than iCANS for all cases except for NH$_3$ case. The total number of shots used are of the same order of magnitude ($10^7$) for all cases with the exception of constant shot sweep with 1000 shots for short version MNIST dataset and with 1000 and 500 shots for full MNIST dataset ($10^8$).

For completeness of analysis, we employ the proposed shot allocation methods for ground state energy estimation problem of different molecules and compare the results obtained with those in previous works. We use constant shot function ($\sim 10^5$ shots), linear function ($4.9*10^4$ shots) and step function ($5.4*10^4$ shots) as shown in Fig. \ref{fig:ground_state_plot} (f) and estimate the ground state energy. We plot the prediction curves along with reference energy (purple line) for BeH$_2$, H$_2$, He$_2^{+}$, LiH, NH$_3$ in Fig. \ref{fig:ground_state_plot} (a)-(e) respectively. We observe that linear function fluctuates a lot compared to constant and step functions, so we say that step function performs the best as it is relatively stable and is also closer to reference ground state energy with lesser number of shots compared to constant function. Comparing predicted ground state energy with reference ground state energy, we observe that in general, there is a difference of $10^{-2}$ Ha $<\Delta $E$<10^{-1}$ Ha for all the molecules, which is slightly worse compared to the CANS related methods. Final prediction of ground state energy depends on the starting molecular Hamiltonian of the molecule provided, which is then optimized using a VQE circuit to obtain the final ground state energy of the molecule. The molecular Hamiltonian is dependent on the geometry of the molecule i.e. the coordinates of atoms in space and bond distance between all the atoms in the molecule. Therefore, a different molecule geometry will give different molecular Hamiltonian and a different prediction for ground state energy. The previous works do not explicitly provide molecular Hamiltonian that are used to find ground state energy. Only \cite{arrasmith2020operator} refer \cite{kandala2017hardware} that provide molecular Hamiltonians of H$_2$, LiH and BeH$_2$ molecules in their supplementary material and even in that some details are missing. Therefore, we obtain the geometry data of all the molecules from PubChem database \cite{wang2008pubchem} to build the molecular Hamiltonian and optimize the VQE to obtain the ground state energy. Note that the molecular geometry data obtained from PubChem database may not be exactly similar to the geometry corresponding the molecular Hamiltonian used in previous works. There may be a slight change, leading to different final ground state energies.\color{black}

Another aspect to be taken into consideration is the effect of MNIST and FMNIST dataset size and quality on the training of the QML model. From the shots sweep of the short version dataset and full dataset in Fig. \ref{fig:sweep_results}, we can observe that the short version dataset has bumps in curves leading to worse performance at times, and the convergence of testing accuracy (90-91\%) is lower compared to full dataset (96-97\%). We observe similar case for FMNIST dataset for its short version. \color{black} These two differences in results can be owed to the size of dataset chosen, and the quality of images chosen. Note that in this work, for both the datasets we randomly select 100 images from each class for training and 25 images from each class for testing, and the results presented for short version are based on the dataset chosen using this method. 
It may not be wise to increase the dataset size since it would imply higher number of total shots required for training. A better approach would be to improve the quality of dataset selected for same size using methods such as, dataset distillation \cite{wang2018dataset} that distills the complete knowledge of a large dataset onto a relatively smaller one. A distilled synthetic dataset from MNIST dataset consisting of only 10 images (1 image per class) can give up to as high as 94\% testing accuracy for a fixed starting initialization and of 100 images (10 images per class) can give up to 80\% testing accuracy for random initialization in very few gradient descent steps (3 epochs). 
\color{black}

\section{Conclusion}
Higher number of shots per iteration could pose a challenge for QML training on hardware due to high training time and access cost. In this paper, we show that higher shots per iteration may not be needed for QML applications. We also found that full dataset provides only 5-6\% improvement in testing accuracy at the cost of one order of magnitude higher number of shots. Therefore, one can optimize the dataset to accelerate training time. We further propose adaptive shot allocation to optimize the number of shots for training time acceleration. The linear shot function requires a lower bound to ensure stable training. For linear shot allocation method, our analysis also showed that the starting shots has a significant impact on testing accuracy compared to the slope of the function. We also used the proposed shot allocation methods on ground state energy estimation problem and conclude that step function is the best in terms of stable energy prediction and efficient shot usage.\color{black}

\bibliographystyle{IEEEtran}
\bibliography{references}

\begin{thebibliography}{10}
\providecommand{\url}[1]{#1}
\csname url@samestyle\endcsname
\providecommand{\newblock}{\relax}
\providecommand{\bibinfo}[2]{#2}
\providecommand{\BIBentrySTDinterwordspacing}{\spaceskip=0pt\relax}
\providecommand{\BIBentryALTinterwordstretchfactor}{4}
\providecommand{\BIBentryALTinterwordspacing}{\spaceskip=\fontdimen2\font plus
\BIBentryALTinterwordstretchfactor\fontdimen3\font minus
  \fontdimen4\font\relax}
\providecommand{\BIBforeignlanguage}[2]{{%
\expandafter\ifx\csname l@#1\endcsname\relax
\typeout{** WARNING: IEEEtran.bst: No hyphenation pattern has been}%
\typeout{** loaded for the language `#1'. Using the pattern for}%
\typeout{** the default language instead.}%
\else
\language=\csname l@#1\endcsname
\fi
#2}}
\providecommand{\BIBdecl}{\relax}
\BIBdecl

\bibitem{gill2022quantum}
S.~S. Gill, A.~Kumar, H.~Singh, M.~Singh, K.~Kaur, M.~Usman, and R.~Buyya,
  ``Quantum computing: A taxonomy, systematic review and future directions,''
  \emph{Software: Practice and Experience}, vol.~52, no.~1, pp. 66--114, 2022.

\bibitem{schuld2019quantum}
M.~Schuld and N.~Killoran, ``Quantum machine learning in feature hilbert
  spaces,'' \emph{Physical review letters}, vol. 122, no.~4, p. 040504, 2019.

\bibitem{biamonte2017quantum}
J.~Biamonte, P.~Wittek, N.~Pancotti, P.~Rebentrost, N.~Wiebe, and S.~Lloyd,
  ``Quantum machine learning,'' \emph{Nature}, vol. 549, no. 7671, pp.
  195--202, 2017.

\bibitem{caro2022generalization}
M.~C. Caro, H.-Y. Huang, M.~Cerezo, K.~Sharma, A.~Sornborger, L.~Cincio, and
  P.~J. Coles, ``Generalization in quantum machine learning from few training
  data,'' \emph{Nature communications}, vol.~13, no.~1, pp. 1--11, 2022.

\bibitem{schuld2020circuit}
M.~Schuld, A.~Bocharov, K.~M. Svore, and N.~Wiebe, ``Circuit-centric quantum
  classifiers,'' \emph{Physical Review A}, vol. 101, no.~3, p. 032308, 2020.

\bibitem{ai2022decompositional}
X.~Ai, Z.~Zhang, L.~Sun, J.~Yan, and E.~Hancock, ``Decompositional quantum
  graph neural network,'' \emph{arXiv preprint arXiv:2201.05158}, 2022.

\bibitem{di2022dawn}
R.~Di~Sipio, J.-H. Huang, S.~Y.-C. Chen, S.~Mangini, and M.~Worring, ``The dawn
  of quantum natural language processing,'' in \emph{ICASSP 2022-2022 IEEE
  International Conference on Acoustics, Speech and Signal Processing
  (ICASSP)}.\hskip 1em plus 0.5em minus 0.4em\relax IEEE, 2022, pp. 8612--8616.

\bibitem{rebentrost2014quantum}
P.~Rebentrost, M.~Mohseni, and S.~Lloyd, ``Quantum support vector machine for
  big data classification,'' \emph{Physical review letters}, vol. 113, no.~13,
  p. 130503, 2014.

\bibitem{li2015experimental}
Z.~Li, X.~Liu, N.~Xu, and J.~Du, ``Experimental realization of a quantum
  support vector machine,'' \emph{Physical review letters}, vol. 114, no.~14,
  p. 140504, 2015.

\bibitem{lloyd2018quantum}
S.~Lloyd and C.~Weedbrook, ``Quantum generative adversarial learning,''
  \emph{Physical review letters}, vol. 121, no.~4, p. 040502, 2018.

\bibitem{chen2022quantum}
S.~Y.-C. Chen, S.~Yoo, and Y.-L.~L. Fang, ``Quantum long short-term memory,''
  in \emph{ICASSP 2022-2022 IEEE International Conference on Acoustics, Speech
  and Signal Processing (ICASSP)}.\hskip 1em plus 0.5em minus 0.4em\relax IEEE,
  2022, pp. 8622--8626.

\bibitem{riste2017demonstration}
D.~Rist{\`e}, M.~P. Da~Silva, C.~A. Ryan, A.~W. Cross, A.~D. C{\'o}rcoles,
  J.~A. Smolin, J.~M. Gambetta, J.~M. Chow, and B.~R. Johnson, ``Demonstration
  of quantum advantage in machine learning,'' \emph{npj Quantum Information},
  vol.~3, no.~1, pp. 1--5, 2017.

\bibitem{ibmcost}
\BIBentryALTinterwordspacing
IBM, ``{IBM} {Q}uantum access plans,'' 2022. [Online]. Available:
  \url{https://www.ibm.com/quantum/access-plans}
\BIBentrySTDinterwordspacing

\bibitem{ionqcost}
\BIBentryALTinterwordspacing
IonQ, ``{I}on{Q} {Q}uantum {C}loud {R}esource {E}stimator,'' 2022. [Online].
  Available:
  \url{https://ionq.com/programs/research-credits/resource-estimator}
\BIBentrySTDinterwordspacing

\bibitem{rigetticost}
\BIBentryALTinterwordspacing
Amazon, ``{A}mazon {B}racket {P}ricing,'' 2022. [Online]. Available:
  \url{https://aws.amazon.com/braket/pricing/}
\BIBentrySTDinterwordspacing

\bibitem{deutsch1989quantum}
D.~E. Deutsch, ``Quantum computational networks,'' \emph{Proceedings of the
  Royal Society of London. A. Mathematical and Physical Sciences}, vol. 425,
  no. 1868, pp. 73--90, 1989.

\bibitem{kubler2020adaptive}
J.~M. K{\"u}bler, A.~Arrasmith, L.~Cincio, and P.~J. Coles, ``An adaptive
  optimizer for measurement-frugal variational algorithms,'' \emph{Quantum},
  vol.~4, p. 263, 2020.

\bibitem{gu2021adaptive}
A.~Gu, A.~Lowe, P.~A. Dub, P.~J. Coles, and A.~Arrasmith, ``Adaptive shot
  allocation for fast convergence in variational quantum algorithms,''
  \emph{arXiv preprint arXiv:2108.10434}, 2021.

\bibitem{arrasmith2020operator}
A.~Arrasmith, L.~Cincio, R.~D. Somma, and P.~J. Coles, ``Operator sampling for
  shot-frugal optimization in variational algorithms,'' \emph{arXiv preprint
  arXiv:2004.06252}, 2020.

\bibitem{balles2016coupling}
L.~Balles, J.~Romero, and P.~Hennig, ``Coupling adaptive batch sizes with
  learning rates,'' \emph{arXiv preprint arXiv:1612.05086}, 2016.

\bibitem{hur2022quantum}
T.~Hur, L.~Kim, and D.~K. Park, ``Quantum convolutional neural network for
  classical data classification,'' \emph{Quantum Machine Intelligence}, vol.~4,
  no.~1, pp. 1--18, 2022.

\bibitem{stronglayers}
\BIBentryALTinterwordspacing
Xanadu, ``qml.{S}trongly{E}ntangling{L}ayers, {P}ennylane {D}ocumentation,''
  2022. [Online]. Available:
  \url{https://pennylane.readthedocs.io/en/latest/code/api/\\pennylane.StronglyEntanglingLayers.html}
\BIBentrySTDinterwordspacing

\bibitem{cai2015entanglement}
X.-D. Cai, D.~Wu, Z.-E. Su, M.-C. Chen, X.-L. Wang, L.~Li, N.-L. Liu, C.-Y. Lu,
  and J.-W. Pan, ``Entanglement-based machine learning on a quantum computer,''
  \emph{Physical review letters}, vol. 114, no.~11, p. 110504, 2015.

\bibitem{xiao2017fashion}
H.~Xiao, K.~Rasul, and R.~Vollgraf, ``Fashion-mnist: a novel image dataset for
  benchmarking machine learning algorithms,'' \emph{arXiv preprint
  arXiv:1708.07747}, 2017.

\bibitem{narkhede2022review}
M.~V. Narkhede, P.~P. Bartakke, and M.~S. Sutaone, ``A review on weight
  initialization strategies for neural networks,'' \emph{Artificial
  intelligence review}, vol.~55, no.~1, pp. 291--322, 2022.

\bibitem{kandala2017hardware}
A.~Kandala, A.~Mezzacapo, K.~Temme, M.~Takita, M.~Brink, J.~M. Chow, and J.~M.
  Gambetta, ``Hardware-efficient variational quantum eigensolver for small
  molecules and quantum magnets,'' \emph{Nature}, vol. 549, no. 7671, pp.
  242--246, 2017.

\bibitem{wang2008pubchem}
Y.~Wang, J.~Xiao, T.~O. Suzek, J.~Zhang, J.~Wang, and S.~H. Bryant, ``Pubchem:
  Integrated platform of small molecules and biological activities,''
  \emph{Annual Reports in Computational Chemistry}, vol.~4, pp. 217--241, 2008.

\bibitem{wang2018dataset}
T.~Wang, J.-Y. Zhu, A.~Torralba, and A.~A. Efros, ``Dataset distillation,''
  \emph{arXiv preprint arXiv:1811.10959}, 2018.

\end{thebibliography}


\begin{IEEEbiography}[{\includegraphics[width=1in,height=1.25in,clip,keepaspectratio]{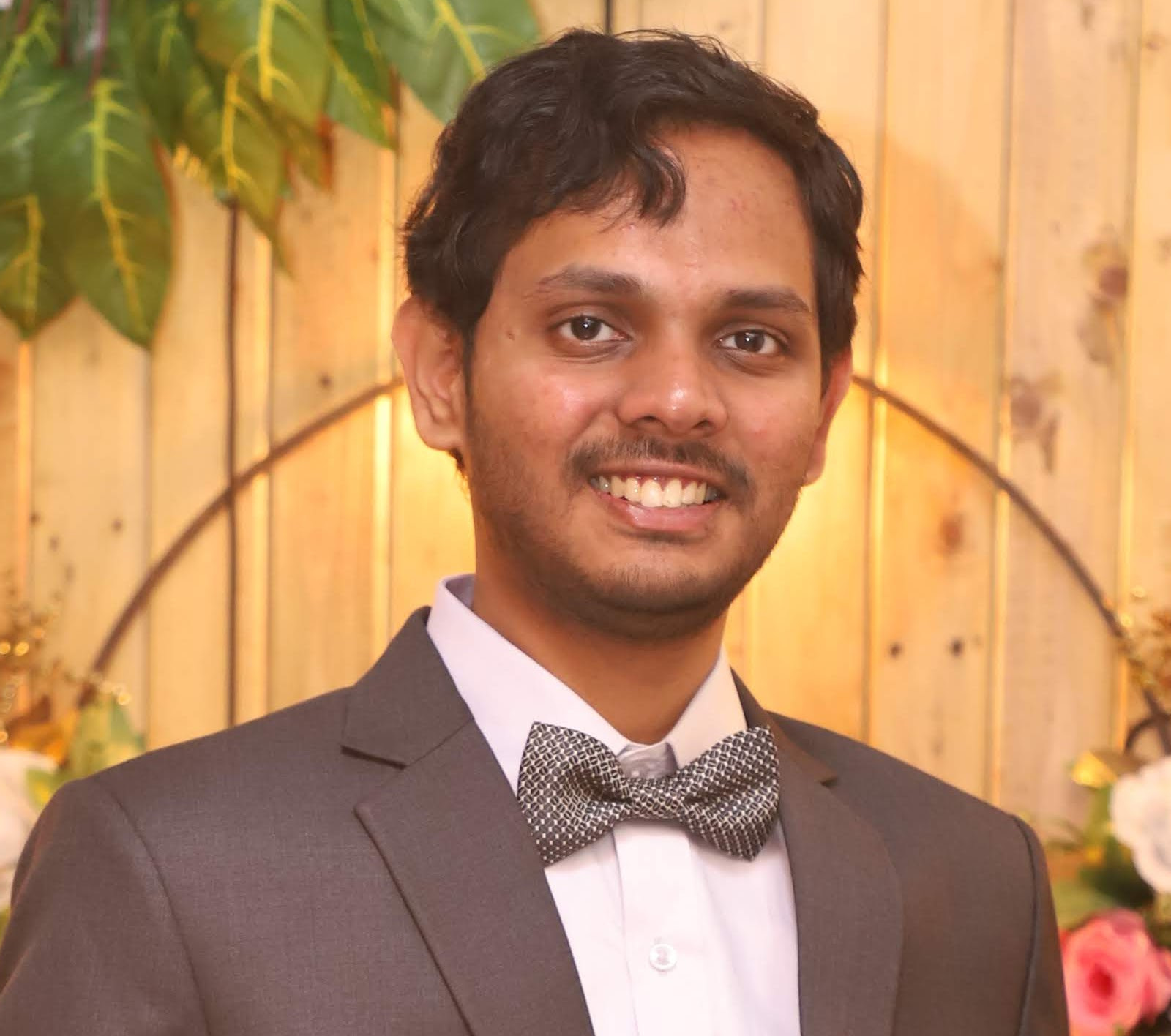}}]{Koustubh Phalak} is a Ph.D. student at the Department of Computer Science and Engineering in Pennsylvania State University. He completed his Bachelors in Electrical and Electronics Engineering from Birla Institute of Technology and Science, Pilani in 2020. He works in the field of emerging technologies, specially quantum computing.
\end{IEEEbiography}


\begin{IEEEbiography}[{\includegraphics[width=1in,height=1.25in,clip,keepaspectratio]{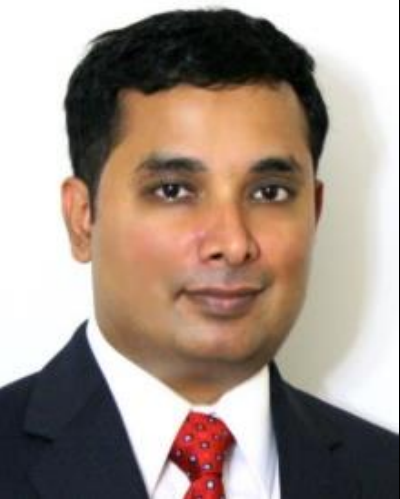}}]{Swaroop Ghosh (SM'13)} received the B.E. (Hons.) from IIT, Roorkee and the Ph.D. degree from Purdue University.
He is an Associate Professor at Pennsylvania State University.
His research interests include quantum computing, emerging memory technologies and hardware security.
			
Dr. Ghosh served as Associate Editor of the IEEE Transactions On Circuits and Systems I and IEEE Transactions On Computer-Aided Design and as Senior Editorial Board member of IEEE Journal of Emerging
Topics on Circuits and Systems (JETCAS). He served as Guest Editor of the IEEE JETCAS and IEEE Transactions On VLSI Systems. He has also served in the technical program committees of more than 25 ACM/IEEE conferences. He served as General Chair, Conference Chair and Program Chair of of ISQED and DAC Ph.D. Forum and track (co)-Chair in DAC, CICC, ISLPED, GLSVLSI, VLSID and ISQED.

Dr. Ghosh is a recipient of Intel Technology and Manufacturing Group Excellence Award, Intel Divisional Award, two Intel Departmental Awards, USF Outstanding Research Achievement Award, College of Engineering Outstanding Research Achievement Award, DARPA Young Faculty Award (YFA), ACM SIGDA Outstanding New Faculty Award, YFA 	Director’s Fellowship, Monkowsky Career Development Award, Lutron Spira Teaching Excellence Award, Dean's Certificate of Excellence and Best Paper Award in American Society of Engineering Education (ASEE). He is a Senior member of the IEEE and the National Academy of Inventors (NAI), Associate member of Sigma Xi and Distinguished Speaker of the Association for Computing Machinery (ACM).

\end{IEEEbiography}

\EOD

\end{document}